\documentclass[letterpaper,twocolumn]{jpsj3}
\usepackage{txfonts}
\usepackage{xspace}
\usepackage{color}

\newcommand{\EF}{$E_\mathrm{F}$\xspace}
\newcommand{\Uf}{$\mathrm{U}~5f$\xspace}
\newcommand{\Udf}{$\mathrm{U}~4d$--$5f$\xspace}
\newcommand{\orb}[2]{$\mathrm{ #1 } ~ #2 $\xspace}
\newcommand{\hn}[1]{$h\nu #1~\mathrm{eV}$\xspace}
\newcommand{\EB}[1]{$E_{\mathrm{B}} #1~\mathrm{eV}$\xspace}
\newcommand{\EBD}[2]{$E_{\mathrm{B}} #1$--$#2~\mathrm{eV}$\xspace}
\newcommand{\UTe}{$\mathrm{UTe_2}$\xspace}
\newcommand{\pnt}[1]{$\mathrm{#1}$\xspace}

\newcommand{\Gm}{$\mathrm{\Gamma}$\xspace}

\title{Electronic Structure of $\mathrm{UTe_2}$ Studied by Photoelectron Spectroscopy}

\author{
Shin-ichi~Fujimori$^1$\thanks{fujimori@spring8.or.jp},
Ikuto~Kawasaki$^1$,
Yukiharu~Takeda$^1$,
Hiroshi~Yamagami$^{1,2}$,
Ai~Nakamura$^3$,
Yoshiya~Homma$^3$,
and Dai~Aoki$^{3}$}

\inst{$^1$Materials Sciences Research Center, Japan Atomic Energy Agency, Sayo, Hyogo 679-5148, Japan \\
$^2$Department of Physics, Faculty of Science, Kyoto Sangyo University, Kyoto 603-8555, Japan \\
$^3$Institute for Materials Research, Tohoku University, Oarai, Ibaraki 311-1313, Japan}%\\
\abst{
The electronic structure of the unconventional superconductor $\mathrm{UTe_2}$ was studied by resonant photoelectron spectroscopy (RPES) and angle-resolved photoelectron spectroscopy (ARPES) with soft X-ray synchrotron radiation.
The partial $\mathrm{U}~5f$ density of states of $\mathrm{UTe_2}$ were imaged by the $\mathrm{U}~4d$--$5f$ RPES and it was found that the $\mathrm{U}~5f$ state has an itinerant character, but there exists an incoherent peak due to the strong electron correlation effects.
Furthermore, an anomalous admixture of the $\mathrm{U}~5f$ states into the $\mathrm{Te}~5p$ bands was observed at a higher binding energy, which cannot be explained by band structure calculations.
On the other hand, the band structure of $\mathrm{UTe_2}$ was obtained by ARPES and its overall band structure were mostly explained by band structure calculations.
These results suggest that the $\mathrm{U}~5f$ states of $\mathrm{UTe_2}$ have itinerant but strongly-correlated nature with enhanced hybridization with the $\mathrm{Te}~5p$ states.
}

\begin{document}
\maketitle

%\section{Introduction}
The unconventional superconductivity in $f$-based materials has attracted much attention over the years. 
Recently, it was discovered that \UTe is one such superconductor with a relatively high transition temperature of $T_\mathrm{SC} = 1.6~\mathrm{K}$ \cite{UTe2_sc}.
Although its transport properties and the nature of its superconductivity have been extensively studied \cite{UTe2_Aoki}, information on its electronic structure is very limited.
In the present study, we have applied \Udf resonant photoelectron spectroscopy (RPES)\cite{U4d5fRPES} and soft x-ray angle-resolved photoelectron spectroscopy (ARPES)\cite{SF_review_JPSJ} to \UTe to unveil its detailed electronic structure.

%\section{Experimental Procedure}
Photoemission experiments were performed on the soft X-ray beamline BL23SU at SPring-8 \cite{BL23SU2}.
The overall energy resolution in the angle-integrated photoelectron spectroscopy (AIPES) experiments at \hn{=800} was about $140~\mathrm{meV}$ and that in the ARPES experiments at \hn{=565-675} was $90-115~\mathrm{meV}$, depending on the photon energies.

The angular resolution of the ARPES experiments was about $\pm 0.15^\circ$, corresponding to a momentum resolution of about $0.065~\mathrm{\AA^{-1}}$ at \hn{=600}.
High quality single crystals of \UTe were grown using a chemical vapor transport method, as previously described in Ref.~\citen{UTe2_sc}.
A clean sample surface was obtained by cleaving the samples perpendicular to the $c$ axis in a ultra-high vacuum chamber.
The positions of the ARPES cuts were determined by assuming a free-electron final state, and the inner potential was taken as $V_{0}=12~\mathrm{eV}$.
For the ARPES spectra, background contributions from elastically scattered photoelectrons due to surface disorder or phonons were subtracted by assuming momentum-independent spectra~\cite{UGe2_UCoGe_ARPES}. 
% ---------- changed ----------
Note that the AIPES and resonant photoemission (RPES) spectra were recorded from the single crystal surface, and they are the integration of the spectra over the rectangular region with the longer side along the $k_x$ direction.
Although the entire Brillouin zone was not covered, the region includes various potions of the Brillouin zone, and the integrated spectra are enough to be compared with calculated pDOS.
% ---------- changed ----------
The vacuum during the course of the measurements was typically $<1.5 \times 10^{-8}~\mathrm{Pa}$, and the sample surfaces were stable for the duration of the measurements ($2-3$ days).
The sample temperature was kept at $20~\mathrm{K}$ for all of the measurements.

%\section{Results and Discussion}
%--------------------------------------------------------------------------------
\begin{figure}[t]
	\centering
	\includegraphics[scale=0.46]{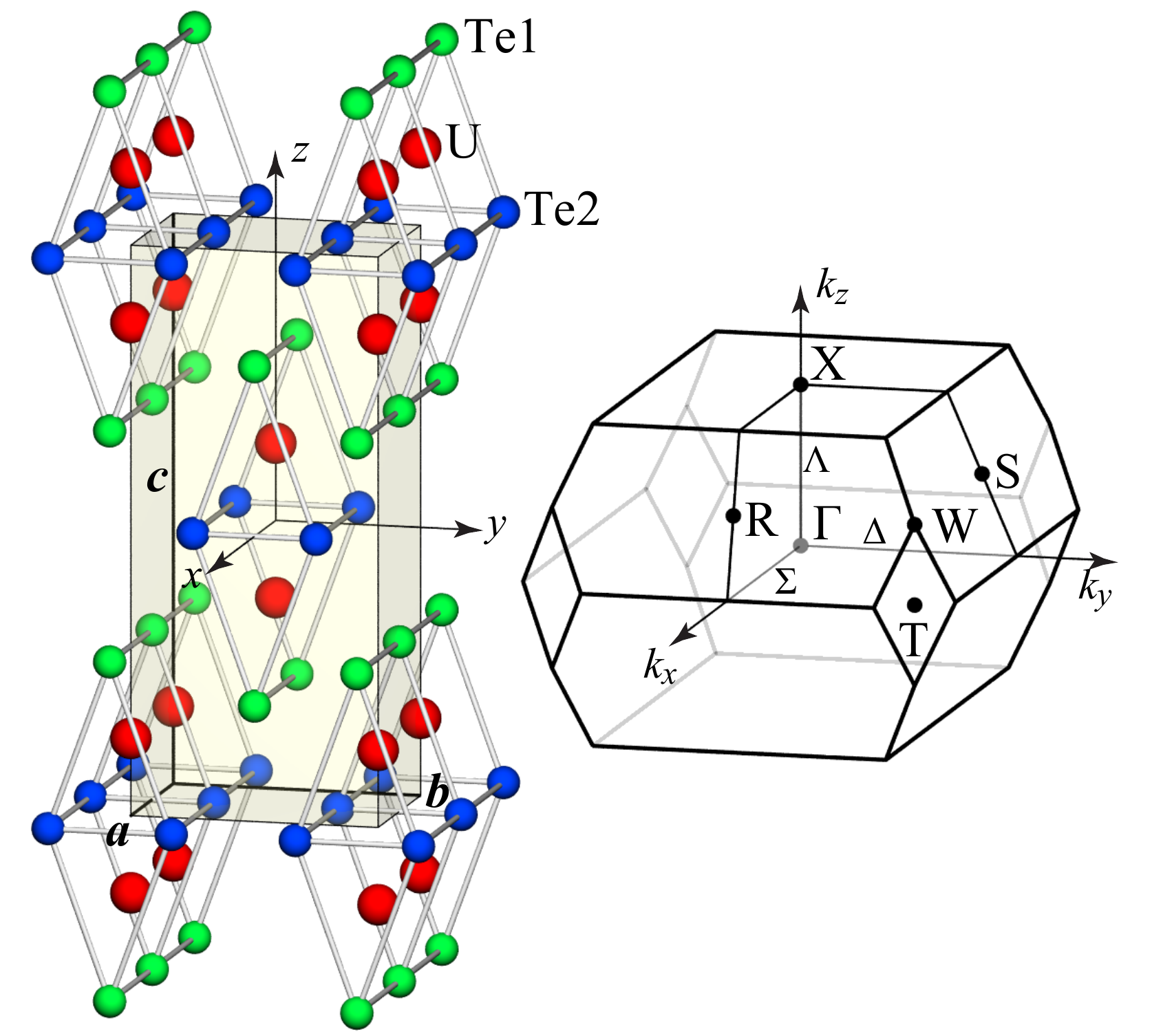}
	\caption{Crystal structure (left) and Brillouin zone (right) of \UTe.
}
	\label{crystal}
\end{figure}
%--------------------------------------------------------------------------------
Figure~\ref{crystal} shows the crystal structure and Brillouin zone of \UTe, which has a body-centered orthorhombic symmetry with $c > b >a$.
%In this case, the corresponding Brillouin zone has a similar shape to that of the body-centered tetragonal one such as the compound with $\mathrm{ThCr_2Si_2}$-type crystal structure, but it has different dimensions along $k_a$ and $k_b$ directions due to its orthorhombic nature.
Note that there are three independent $\Gamma$--\pnt{X} high-symmetry lines, namely along the \Gm--(\pnt{\Sigma})--\pnt{X}, \Gm--(\pnt{\Delta})--\pnt{X}, and \Gm--(\pnt{\Lambda})--\pnt{X} lines corresponding to the $k_x$, $k_y$, and $k_z$ directions, respectively.
In the present study, the ARPES scans were carried out by changing the electron detection angle along the $a$ axis ($k_x$ direction) and the incident photon energy, which correspond to a two dimensional scan within the $k_x$--$k_z$ plane in the momentum space.

%--------------------------------------------------------------------------------
\begin{figure}[t]
	\centering
	\includegraphics[scale=0.46]{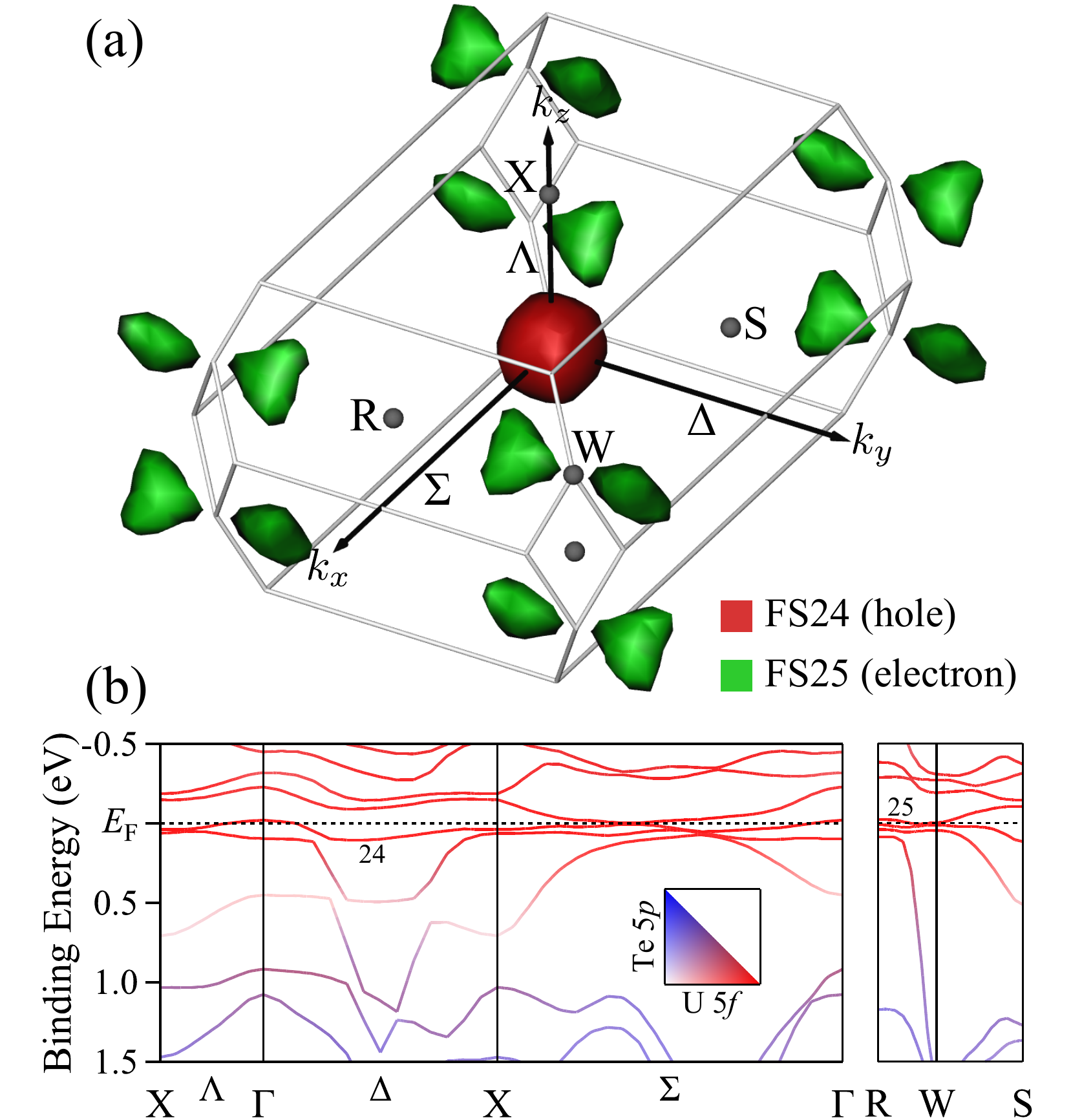}
	\caption{Calculated Fermi surface and band structure of \UTe.
	(a) Calculated Fermi surface and (b) calculated band structure in the vicinity of \EF.
	The color coding of each band represents the contributions from the 	\Uf and \orb{Te}{5p} states. 
}
	\label{FS_calc}
\end{figure}
%--------------------------------------------------------------------------------
Band structure calculations for \UTe were carried out, treating all of the \Uf electrons as being itinerant using the relativistic linear augmented plane wave method \cite{Yamagami} within a local density approximation \cite{LDA}.
Figure~\ref{FS_calc} (a) shows the calculated Fermi surfaces of \UTe.
%In Fig.~\ref{FS_calc}, we show the calculated Fermi surface and band structure.
The calculations predict that \UTe is a semimetal, consistent with the metallic nature of \UTe.
Band 24 forms a spherical hole pocket around the \Gm point, while band 25 forms a twisted pair of heart-like electron pockets along the \pnt{R}--\pnt{W}--\pnt{R} high-symmetry line.
% ---------- changed ----------
The calculated band structure is shown in Fig.~\ref{FS_calc} (b), in which the contributions from the \Uf and \orb{Te}{5p} states are indicated by the color coding of each band.
% ---------- changed ----------
The calculations predicted some less-dispersive bands with large contribution from the \Uf states in the very vicinity of the Fermi energy (\EF), forming very shallow electron or hole pockets.
For example, the top of band 24 and the bottom of band 25 are about $20~\mathrm{meV}$ above and $8~\mathrm{meV}$ below \EF, respectively.
Thus, a tiny change in the Fermi energy drastically transforms the shape of the Fermi surface.
This situation is very similar to the case of ferromagnetic heavy Fermion superconductors such as $\mathrm{URhGe}$ \cite{URhGe_ARPES} and $\mathrm{UCoGe}$ \cite{UGe2_UCoGe_ARPES}.

\begin{figure}
	\centering
	\includegraphics[scale=0.46]{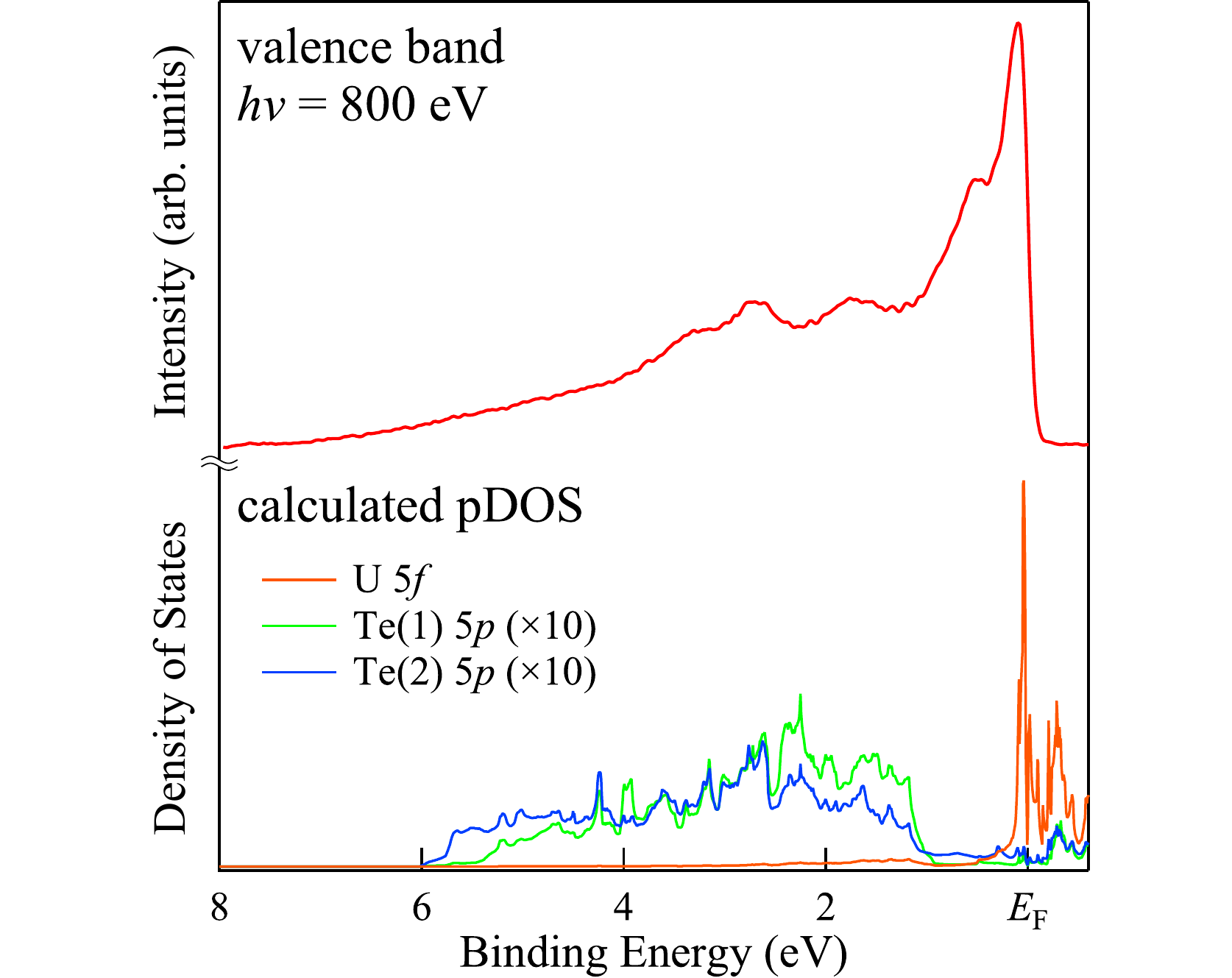}
	\caption{Valence band spectrum of \UTe measured at \hn{=800}, and the theoretical \Uf and \orb{Te}{5p} pDOS obtained from the band structure calculations.
}
	\label{AIPES}
\end{figure}

Figure~\ref{AIPES} shows the AIPES spectrum of \UTe measured at \hn{=800} and the partial density of states (pDOS) obtained from the band structure calculations.
The spectrum has a sharp peak structure just below the \EF and long tail toward higher binding energies.
According to the calculated photoionization cross sections of the atomic orbitals \cite{Atomic}, the contribution from the \Uf states is dominant in this photon energy range.
The second largest contribution is that from the \orb{Te}{5p} state, the cross-section of which is about 15 \% that of the \Uf state.
Thus, the \orb{Te}{5p} state should show a tiny contribution to the spectrum.
The \Uf and \orb{Te}{5p} pDOS obtained from the band structure calculations are also presented in the bottom of the figure.
Note that the \orb{Te}{5p} pDOS are ten times enhanced in the figure.
The sharp peak structure just below \EF in the experimental spectrum seems to correspond to the calculated \Uf pDOS, but the spectrum is much broader than the calculated \Uf pDOS.
%A comparison between the experimental spectrum and calculated pDOS suggests that the sharp peak just below \EF corresponds to the contribution from \Uf states.
%It is understood that the sharp peak just below the Fermi energy corresponds to the contribution from \Uf states.
%On the other hand, the experimental spectrum is much broader than the calculated \Uf pDOS.
A shoulder structure exists at \EB{\sim 0.5} and the spectral intensities at \EBD{=1}{6}, which should be contributions from the \orb{Te}{5p} states, are enhanced in the spectrum.

%--------------------------------------------------------------------------------
%--------------------------------------------------------------------------------
%--------------------------------------------------------------------------------
\begin{figure*}[t]
	\centering
	\includegraphics[scale=0.46]{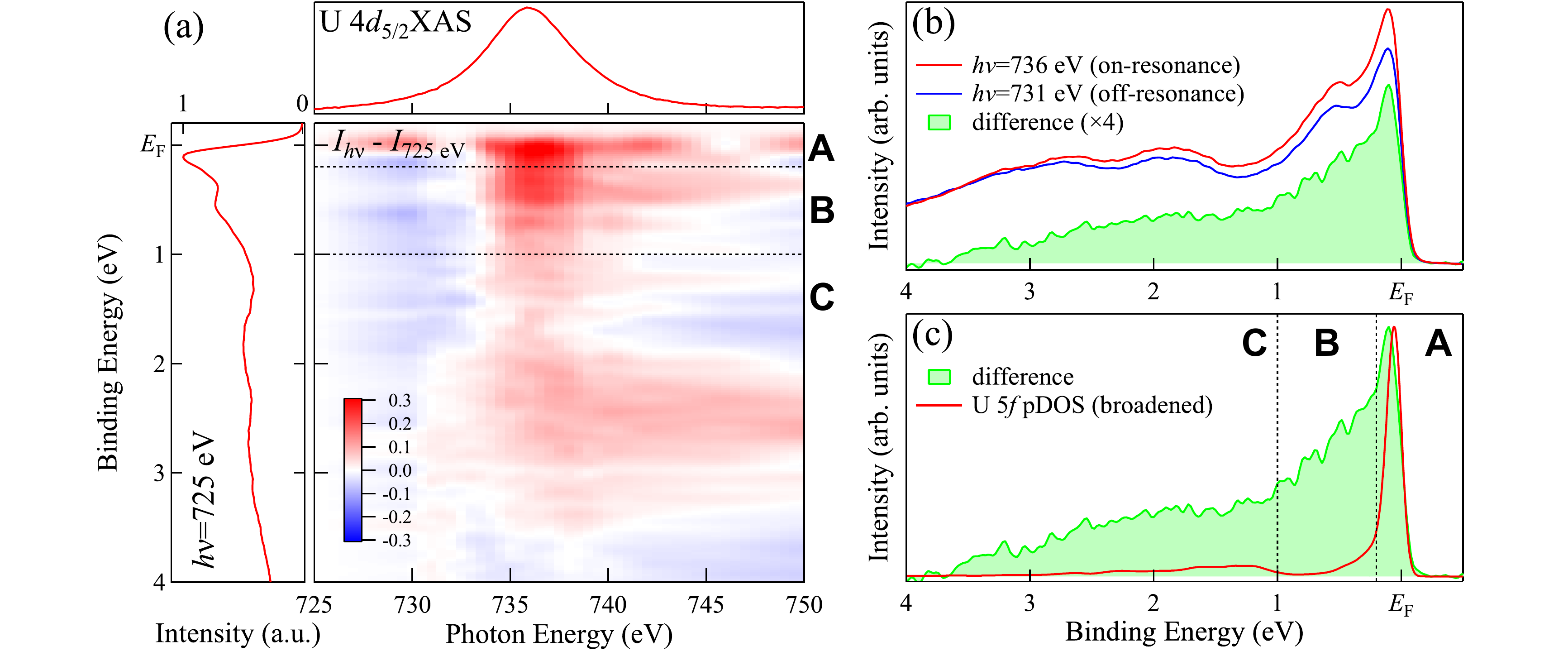}
	\caption{RPES spectra of \UTe.
		(a) Density plot of RPES spectra together with the \orb{U}{4d_{5/2}} XAS spectrum.
		(b) On- and off-resonance spectra measured at $h \nu = 736$ and $731~\mathrm{eV}$, respectively, and the corresponding difference spectrum.
		(c) Comparison of the difference spectrum and the calculated \Uf pDOS.
}
	\label{RPES}
\end{figure*}
%--------------------------------------------------------------------------------
%--------------------------------------------------------------------------------
\begin{figure*}[t]
	\centering
	\includegraphics[scale=0.46]{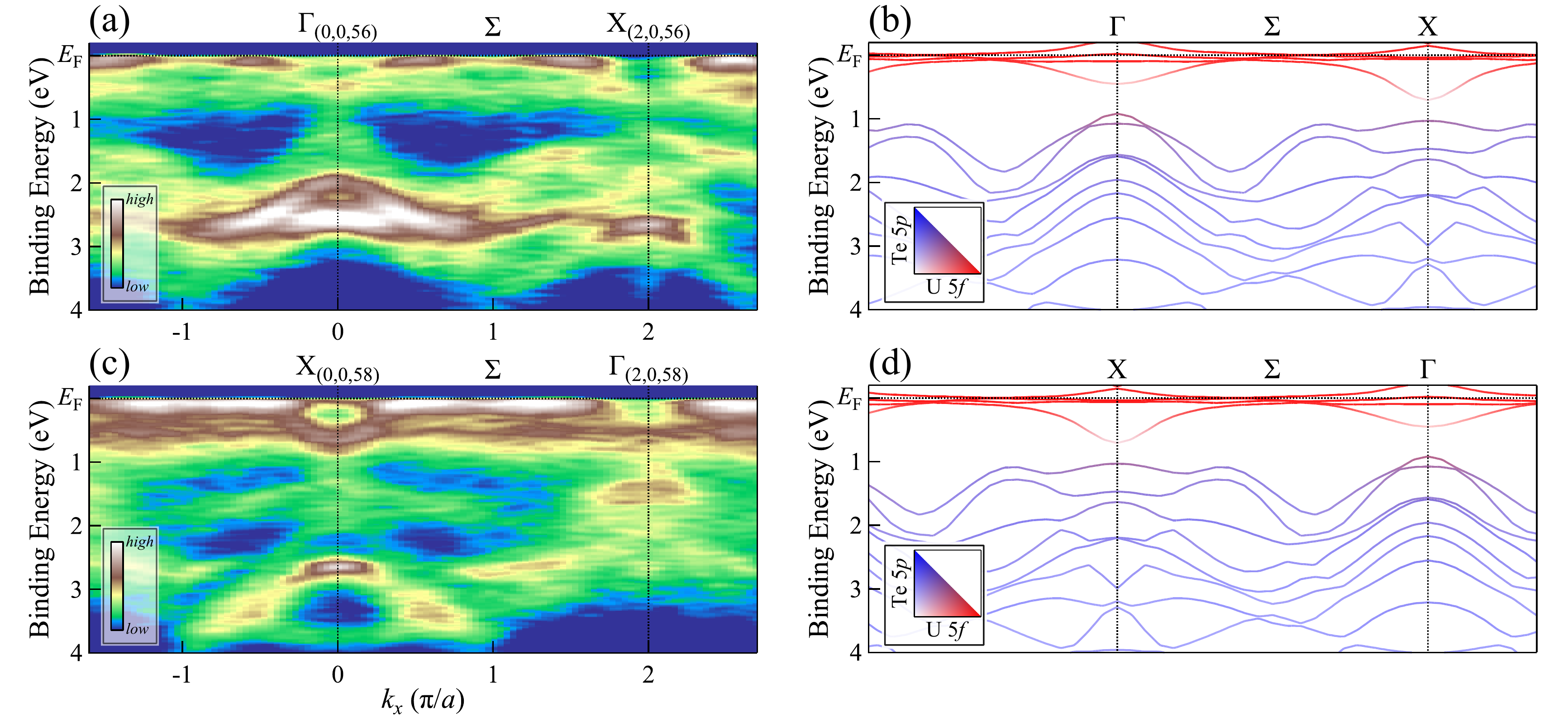}
	\caption{ARPES spectra of \UTe, together with the corresponding results of the band structure calculations.
	(a) ARPES spectra measured along the $\mathrm{\Gamma}_{(0,0,56)}$--$\mathrm{(\Sigma)}$--$\mathrm{X}_{(2,0,56)}$ high-symmetry line.
%	The denotation of the point represents the location of the momentum space in the units of $(\pi/a)$, $(\pi/b)$, and $(\pi/c)$ for $k_x$, $k_y$, and $k_z$ directions, respectively (See Fig.~\ref{ARPES_hn_FS} (a)).	
	(b) The corresponding calculated band structure.
	The color coding represents the contribution from the \Uf states.
	(c) ARPES spectra measured along the $\mathrm{X}_{(0,0,58)}$--$\mathrm{(\Sigma)}$--$\mathrm{\Gamma}_{(2,0,58)}$ high-symmetry line.
	(d) Corresponding  calculated band structure and the simulated ARPES spectra.
}
	\label{ARPES_hn_band}
\end{figure*}
%--------------------------------------------------------------------------------
To further clarify the energy distribution of the \Uf state in the valence band, \Udf RPES \cite{U4d5fRPES} was applied to \UTe, and the results are summarized in Fig.~\ref{RPES}.
%Figure~\ref{RPES} shows the \Udf RPES spectra and the \orb{U}{4d_{5/2}} X-ray absorption spectroscopy (XAS) spectrum of \UTe.
The top and left panels in Fig.~\ref{RPES} (a) represent the \orb{U}{4d_{5/2}} X-ray absorption spectroscopy (XAS) spectrum and photoemission spectrum measured at \hn{=725}, respectively.
The XAS spectrum has a maximum at \hn{= 736} and the spectrum measured at \hn{=725} corresponds to the complete off-resonant condition.
The density plot in the center of Fig.~\ref{RPES} (a) represents the difference between the spectrum measured at \hn{=725} and that measured at each photon energy, and all of the spectra were normalized to the maximum of the spectrum measured at \hn{=725}.
%The horizontal and vertical axes are the incident photon energy and the binding energy, respectively. 

As the photon energy approaches the \orb{U}{4d_{5/2}} absorption edge, the spectra exhibit a clear enhancement, with noticeable dependence on the binding energy.
This enhancement was categorize into three binding-energy regions, A, B, and C as shown in Fig.~\ref{RPES} (a).
In binding energy region A (\EB{\lesssim 0.2}), the sharp peak just below \EF exhibits a strong enhancement at the absorption edge.
This is the common structure of itinerant \Uf compounds, which originates from itinerant quasi-particle bands.
%The enhancement of the signal is about 30 \%, which is somewhat higher that the cases of other uranium compounds (15-25 \%) \cite{U4d5fRPES}.
In the binding energy region B ($0.2 \lesssim E_\mathrm{B} \lesssim 1.0$), there are other independent enhancements.
The nature of these enhancements is very similar to the one observed in the same binding energy region of the RPES spectra of $\mathrm{UPd_{2}Al_{2}}$\cite{U4d5fRPES} which is due to the contribution from the incoherent peak originating from electron correlation effects.
The similarities in the shape and binding energy suggest that there is an incoherent peak in \UTe.
Surprisingly, in binding energy region C ($1.0 \lesssim E_\mathrm{B} \lesssim 3.5$), where the contributions from the \orb{Te}{5p} states are dominant in the band structure calculations, the RPES spectra exhibit weak enhancement.
This means that there is a finite contribution from the \Uf states in this binding energy region. 

% ---------- changed ----------
Figure.~\ref{RPES} (b) shows the on-resonance (\hn{=736}) and off-resonance (\hn{=731}) spectra, along with the corresponding difference spectrum.
% ---------- changed ----------
The shape of the difference spectrum is similar to those of the valence band spectra of itinerant uranium compounds such as $\mathrm{UAl}_3$ \cite{U4d5fRPES}, but its tail persists until a much higher binding energy.
The difference spectrum is compared with the calculated \Uf pDOS in Fig.~\ref{RPES} (c).
The red curve represents the \Uf pDOS obtained from the band structure calculations, which were multiplied by the Fermi--Dirac function and broadened by the instrumental energy resolution to simulate the experimental \Uf difference spectrum.
The difference spectrum is much broader than the calculated \Uf pDOS.
As can be seen in the resonant enhancement shown in Fig~\ref{RPES} (a), there are two different contributions to the broadening.
In region B, there is the incoherent peak with a dominant contribution from the \Uf state.
In region C, the finite \Uf contribution is admixed into the \orb{Te}{5p} band, which is not predicted by the band structure calculations.
These results suggest that the \Uf states in \UTe are itinerant in nature, but there are some deviations from the simple itinerant description of the \Uf states.

%--------------------------------------------------------------------------------
\begin{figure}
	\centering
	\includegraphics[scale=0.46]{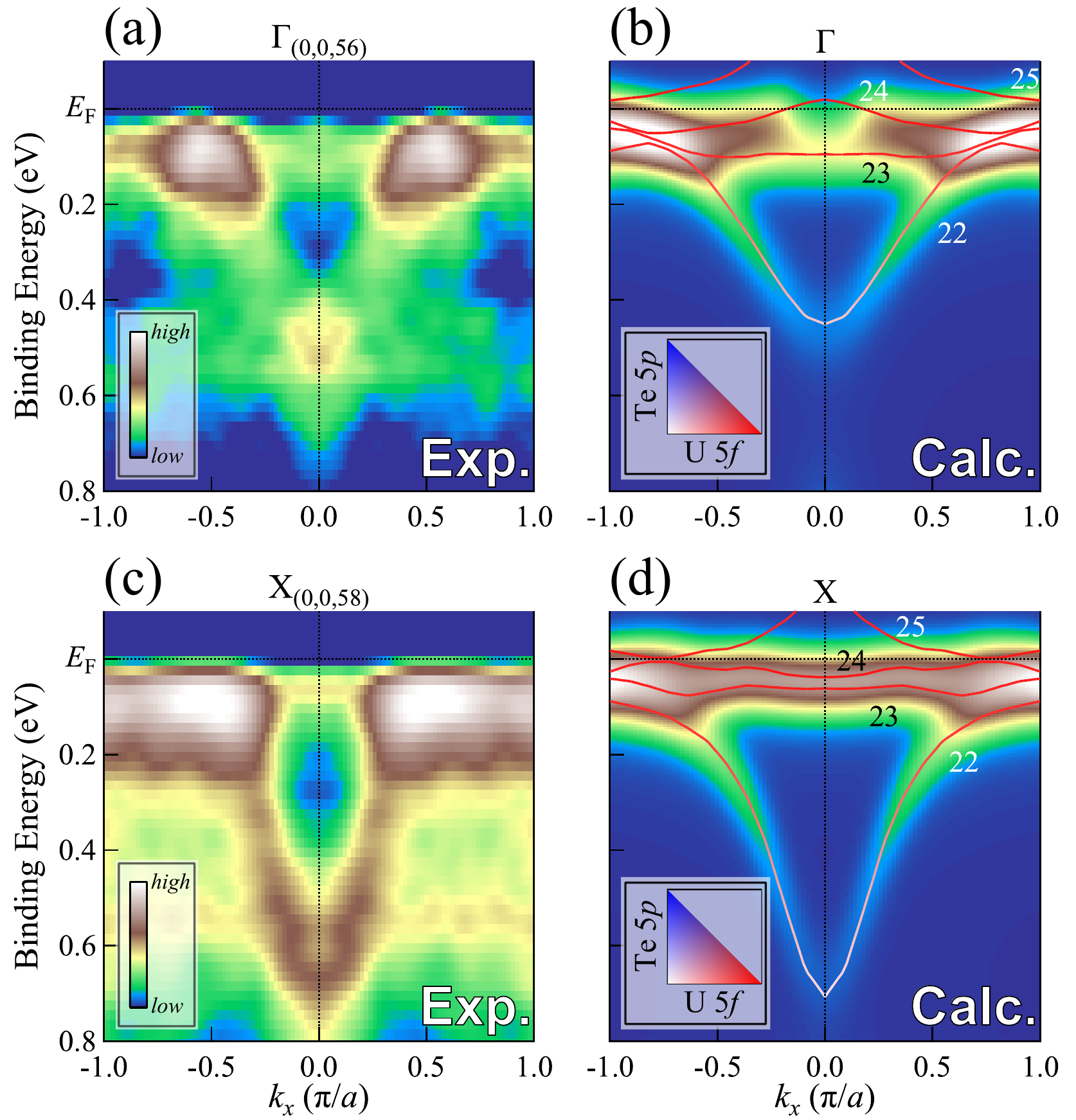}
	\caption{Magnified ARPES spectra and the corresponding results of the band structure calculations.
	(a) ARPES spectra recorded around the $\mathrm{\Gamma}_{(0,0,56)}$ point.
	(b) Simulation of the ARPES spectra based on the band structure calculations.
	(c) Same as (a) but around the $\mathrm{X}_{(0,0,58)}$ point.
	(d) Same as (b) but around the \pnt{X} point.
}
	\label{ARPES_hn_EF}
\end{figure}
\begin{figure*}[t]
	\centering
	\includegraphics[scale=0.46]{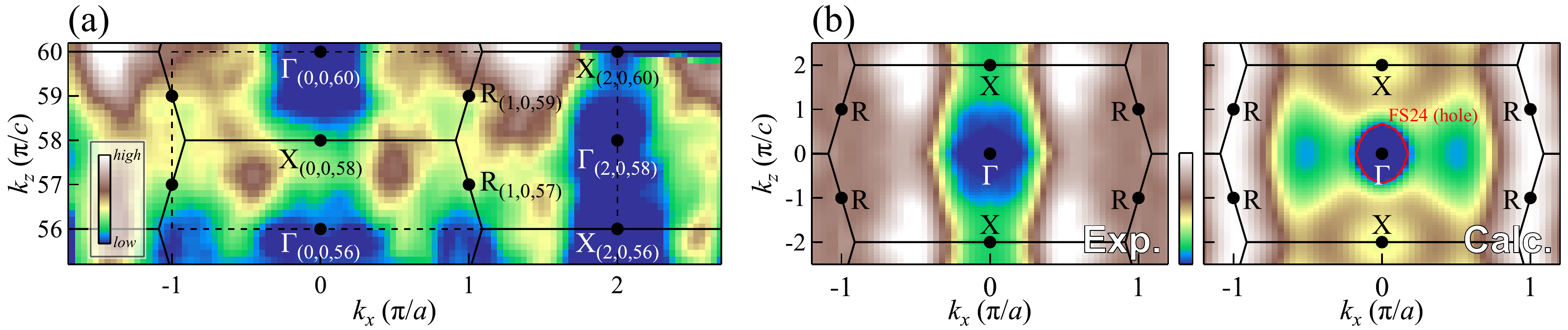}
	\caption{Fermi surfaces of \UTe obtained by photon energy scanning of the ARPES measurements, together with the corresponding band structure calculation results.
	(a) Fermi surfaces of \UTe obtained by integrating the ARPES spectra measured at $h\nu =655$--$745~\mathrm{eV}$ over $\mathrm{100~meV}$ at \EF.
	(b) Symmetrized and folded Fermi surface map within the $k_x$--$k_z$ plane and the simulated Fermi surface map with the calculated Fermi surface.
}
	\label{ARPES_hn_FS}
\end{figure*}

%--------------------------------------------------------------------------------
%Next, we discuss the ARPES spectra of \UTe.% measured within the $k_x$--$k_z$ plane.
Figure~\ref{ARPES_hn_band} (a) shows the ARPES spectra of \UTe measured along the $\mathrm{\Gamma}_{(0,0,56)}$--$\mathrm{(\Sigma)}$--$\mathrm{X}_{(2,0,56)}$ high-symmetry line.
Here the denotation of the point represents the location of the momentum space in the units of $\pi/a=0.7550~\mathrm{[\AA^{-1}]}$, $0.5131~\mathrm{[\AA^{-1}]}$, and $\pi/c=0.2251~\mathrm{[\AA^{-1}]}$ for the $k_x$, $k_y$, and $k_z$ directions, respectively (See Fig.~\ref{ARPES_hn_FS} (a)).
The spectra correspond to the ARPES scan with \hn{\sim 580}.
%The ARPES spectra exhibit clear energy dispersions.
The weakly dispersive features in the vicinity of \EF are quasi-particle bands with dominant contributions from the \Uf states.
Their intensities have momentum dependence in the very vicinity of \EF, suggesting that the quasi-particle bands might form Fermi surfaces.
Meanwhile, the dispersive bands at higher binding energies (\EB{\gtrsim 1}) are bands with dominant contributions from the \orb{Te}{5p} states, although there are finite admixtures from the \Uf states, as can be seen form the RPES spectra.
Figure~\ref{ARPES_hn_band} (b) shows the corresponding calculated band structure.
The color coding of the bands represents the projection of the contributions from the \Uf and \orb{Te}{5p} states.
Generally, the bands in \EB{\lesssim 1} and \EB{\gtrsim 1} have dominant contributions from the \Uf state and the \orb{Te}{5p} states, respectively.
%The bands in \EB{\lesssim 1} have a dominant contribution from \Uf state while the bands in \EB{\gtrsim 1} have a dominant contribution from the \orb{Te}{5p} states.
Although each of the band is not clearly resolved in the experimental ARPES spectra, their overall structure can be mostly explained by the calculated bands.
For example, the inverted parabolic dispersions around the \Gm and \pnt{X} points in the binding energy of \EB{\gtrsim 1} correspond well to the calculated bands located at the same binding energies.
In addition, the less-dispersive features just below \EF also coincide with the calculated bands that have dominant contributions from the \Uf state, the details of which are discussed later.
Figures~\ref{ARPES_hn_band} (c) and (d) show the ARPES spectra measured along the $\mathrm{X}_{(0,0,58)}$--$\mathrm{(\Sigma)}$--$\mathrm{\Gamma}_{(2,0,58)}$ high-symmetry line and the corresponding calculated band structure, respectively.
The spectra correspond to the ARPES scan with \hn{\sim 620}.
The essential structure of the spectra is very similar to that measured along the $\mathrm{\Gamma}_{(0,0,56)}$--$\mathrm{(\Sigma)}$--$\mathrm{X}_{(2,0,56)}$ high-symmetry line, but the intensities of the dispersions are different due to the momentum matrix element effect.
The different appearance of the spectra makes it further possible to follow the energy dispersions.
In particular, the intensity around the \pnt{X} point is more enhanced.
As in the case of the $\mathrm{\Gamma}_{(0,0,56)}$--$\mathrm{(\Sigma)}$--$\mathrm{X}_{(2,0,56)}$ high-symmetry line shown in Fig.~\ref{ARPES_hn_band} (a), the overall structure of the ARPES spectra can be mostly explained by the calculated bands.
%there are good correspondences between experiment and the calculation.
%On the other hand, the intensity of the non-dispersive feature at \EB{\sim 0.5} is much enhanced, and again there is no corresponding band in the calculation.

%--------------------------------------------------------------------------------
%--------------------------------------------------------------------------------
% ---------- changed ----------
Figure~\ref{ARPES_hn_EF} shows magnified regions of the ARPES spectra in the vicinity of \EF and the simulation of the ARPES spectra based on the band structure calculations.
% ---------- changed ----------
In the simulation, the following factors were taken into account: the broadening along the $k_z$ direction due to the finite escape depth of the photoelectrons, the lifetime broadening of the photohole, the photoionization cross sections of the orbitals, and the energy resolution and angular resolution of the photoelectron analyzer, the details of which are described in Ref.~\citen{UN_ARPES}.
The experimental ARPES spectra were symmetrized relative to the high-symmetry points (\Gm or \pnt{X}) to obtain better statistics. 
Figure~\ref{ARPES_hn_EF} (a) and (b) show the ARPES spectra and the results of the simulation around the \Gm point, respectively.
The experimentally-observed parabolic energy dispersion corresponds well to that of the calculated band 22.
Meanwhile, the correspondences of the calculated bands 23--25 to the experimental spectra were not clear due to their very narrow dispersive natures, but the intensity maps show some similarity between the experimental data and calculations.
%there should be analogue bands in the corresponding energy region since there are some similarities in the spectral intensities in the very vicinity of \EF.
On the other hand, the non-dispersive feature at \EB{= 0.5-0.6} does not have a corresponding band and is considered to be the contribution from the incoherent peak observed in the RPES spectra.
%On the other hand, the non-dispersive feature at \EB{= 0.5-0.6} does not have a corresponding band and is considered to arise as a result of the contribution from the incoherent peak observed in the RPES spectra.
%there is almost non-dispersive feature at \EB{\sim 0.5-0.6}, but there is no corresponding band in the calculation.
%This is the contribution from the incoherent peak observed in the RPES spectra.
The situation is very similar to the case around the \pnt{X} point as shown in Figs.~\ref{ARPES_hn_EF} (c) and (d).
Band 22 is in good agreement well with the experimental spectra and the intensity maps in the vicinity of \EF are similar in both the experimental data and the calculations, although each band was not resolved.
Besides this, the non-dispersive feature at \EB{\sim 0.5-0.6} does not have a corresponding band in the calculations as in the case around the \Gm point.
Accordingly, it was found that the electronic structure in the vicinity of \EF is similar to that of the band structure calculations although the details of the Fermi surface topology were not resolved experimentally.
In addition, an incoherent peak was observed at \EB{\sim 0.5-0.6} in the ARPES spectra, which cannot be explained by the band structure calculations. 
%the situation is the same to the case of the \Gm point.%, implying it is the incoherent peak.
%There are shallow parabolic energy dispersions around the \Gm and the \pnt{X} points, and their essential structures agree with the calculated energy bands.
%On the other hand, there is almost non-dispersive feature at \EB{\sim 0.5}, but there is no corresponding band in the calculation.
%Its energy position is identical to that of incoherent peak observed in the RPES spectra.
% ---------- changed ----------
Note that the renormalization of bands should be significant only in the very vicinity of \EF where the contributions from the \Uf states are enhanced.
% ---------- changed ----------

% ---------- changed ----------
To further understand the electronic structure in the vicinity of \EF, we discuss the Fermi surface map of \UTe.
%the Fermi surface map of \UTe is shown in Fig~\ref{ARPES_hn_FS}.
% ---------- changed ----------
Figure~\ref{ARPES_hn_FS} (a) shows the Fermi surface map within the $k_x$--$k_z$ plane obtained by integrating the ARPES spectra over $100~\mathrm{meV}$ at \EF.
Note that the locations with the same symmetry in the Brillouin zone but different values of $k_x$ and $k_z$ have different profiles.
%Particularly, the intensities around the \pnt{X} point are considerably different in different locations in the momentum space. 
The different appearances are due to the momentum matrix element effect as has already been mentioned.
To avoid this effect as much as possible, the Fermi surface map was symmetrized and folded within the momentum region within $k_x =-1$ -- $2 (\pi/a)$ and $k_z =56$ -- $60 (\pi/c)$ as indicates by the dashed lines in Fig.~\ref{ARPES_hn_FS} (a).
Figure~\ref{ARPES_hn_FS} (b) shows a comparison between the symmetrized and folded Fermi surface map and the simulated Fermi surface map based on the band structure calculations.
Within this high-symmetry plane, the calculations predict a hole pocket formed by band 24.
Although the Fermi surface maps have different intensity distributions in the experimental data and calculations, some similarities between them can be recognized.
In particular, the intensity around the \Gm point is reduced in both the experimental data and the calculations, and their shapes are similar to each other.
In addition, the intensities around the \pnt{R} point are enhanced in both the experimental data and the calculations although their shapes are somewhat different. 
These similarities in the Fermi surface map as well as the bands in the vicinity of \EF between the experimental data and calculations suggest that the essential structure of the Fermi surface should not be so different from the band structure calculation although their topologies were not directly determined. 

In summary, \Udf RPES and ARPES were applied to \UTe.% and it was revealed that \Uf states have itinerant but strongly correlated nature.
The overall band structure of \UTe can be described by band structure calculations, although there is the non-dispersive incoherent band at \EB{=0.5-0.6} originated from the strong electron correlation effects.
The topology of the Fermi surface was not resolved experimentally due to the very narrow nature of the bands in the vicinity of \EF, but similar spectral profiles were observed in the experimental data and the band structure calculation.
% ---------- changed ----------
Thus, the electronic structure of \UTe can be described by taking into account the electron correlation effect to the itinerant \Uf model, and the superconductivity in \UTe is mediated by the heavy quasi-particles. 
% ---------- changed ----------
%On the other hand, there is the incoherent peak in \Uf difference spectrum, suggesting that 
On the other hand, the \Uf states are mixed with the \orb{Te}{5p} bands distributed at deeper binding energies, the behavior of which could not be explained by the band structure calculations.
The origin of the anomalous admixture is not understood at present.

\begin{acknowledgment}
The experiment was performed under Proposal Nos 2019A3811 at SPring-8 BL23SU.
The present work was financially supported by JSPS KAKENHI Grant Numbers JP15H05882, JP15H05884, JP15H05745, JP15K21732, JP16H01084, JP16H04006, JP18K03553, and JP19H00646.
\end{acknowledgment}

\bibliographystyle{jpsj}
\bibliography{UTe2}

\end{document}